\documentclass[twocolumn,pra,longbibliography,nobalancelastpage,aps,raggedbottom, superscriptaddress]{revtex4-2}
\usepackage{amsmath,amsfonts,amssymb,scalerel,natbib}
\usepackage{graphicx,hyperref,epstopdf,xcolor,tikz}
\usepackage{array,mathptmx,textcomp,braket,times}
\usepackage{tabularx,makecell,multirow}
\usepackage[utf8]{inputenc}
\usepackage[T1]{fontenc}
\usepackage{relsize}

\definecolor{lime}{HTML}{A6CE39}
\DeclareRobustCommand{\orcidicon}
{
	\begin{tikzpicture} 
	\draw[lime, fill=lime] (0,0) circle [radius=0.15] node[white] {{\fontfamily{qag}\selectfont \tiny ID}};
	\draw[white, fill=white] (-0.0625,0.095) 	circle [radius=0.007];
	\end{tikzpicture}
	\hspace{-2.2mm}
}
\newcommand\orcidID[1]{\href{https://orcid.org/#1}{\orcidicon}}

\newcommand{\be}{\begin {equation}}
\newcommand{\ee}{\end {equation}}
\newcommand{\beqa}{\begin {eqnarray}}
\newcommand{\eeqa}{\end {eqnarray}}
\newcommand{\mb}{\mathbf}
\newcommand{\Sch}{Schr\"odinger }

\hypersetup{colorlinks,citecolor=blue,filecolor=black,linkcolor=blue,urlcolor=blue}

\begin{document}
\title{Backward-scattering photoelectron holography of a polar molecule driven by a sub-cycle laser pulse}

\author{Rambabu Rajpoot\orcidID{0000-0002-2196-6133}}
\email[E-mail: ]{rambabu.rajpoot@riken.jp}
\affiliation{Extreme Photonics Research Team, RIKEN Center for Advanced Photonics, RIKEN, 2-1 Hirosawa, Wako, Saitama 351-0198, Japan}

\author{Eiji J. Takahashi\orcidID{0000-0003-1672-1171}}
\email[Corresponding author: ]{ejtak@riken.jp}
\affiliation{Extreme Photonics Research Team, RIKEN Center for Advanced Photonics, RIKEN, 2-1 Hirosawa, Wako, Saitama 351-0198, Japan}

\date{\today}
\begin{abstract}
We investigate photoelectron holographic interference in the polar CO molecule driven by a sub-cycle laser pulse using time-dependent \Sch equation simulations and a classical-trajectory-based analysis. The sub-cycle temporal confinement of the driving field enables the forward-scattering holographic (FSH) and backward-scattering holographic (BSH) interference structures in the photoelectron momentum distributions (PMDs) to be clearly distinguished. The FSH fringes exhibit a strong dependence on molecular orientation, which is associated with the orientation-dependent transverse displacement of the ionizing electron wave packet (EWP). Furthermore, the heteronuclear nature of CO results in different FSH fringe shifts for ionization occurring predominantly from the C and O sides of the molecule. In contrast, the curved interference structures in the negative parallel-momentum region are identified as BSH interference arising from the coherent superposition of direct and backward-scattered EWPs. The BSH fringe positions show weak dependence on molecular orientation and remain clearly visible after orientation averaging.  We further show that BSH visibility depends on the carrier-envelope phase (CEP) and that the fringes remain discernible after CEP averaging.  Moreover, the BSH interference pattern exhibits a systematic dependence on the internuclear separation, demonstrating its sensitivity to molecular geometry. These results highlight the potential of sub-cycle pulse enabled backward-scattering photoelectron holography for probing molecular structure and demonstrate its robustness against molecular-orientation and CEP averaging.
\end{abstract}

\maketitle

\section{Introduction}
\label{sec1}
Strong-field photoelectron holography (SPH) provides a powerful approach for probing ultrafast electron dynamics and retrieving structural information in atoms and molecules on sub-laser-cycle timescales and with sub-angstrom spatial resolution~\cite{Spanner2004_JPhysB, Barton1988_PRL, Huismans2011_Science, Bian2011_PRA, Huismans2012_PhysRevLett, Haertelt2016_PRL, Shvetsov2018_PRA, Faria2020_RepProgPhys}. Following strong-field tunneling ionization, a fraction of the ionized EWP can be driven back toward the parent ion and undergo scattering, while another portion can be driven directly to the detector without rescattering~\cite{Krause1992_PRL,Corkum1993_PRL,Schafer1993_PRL}. The rescattered EWP acts as a signal wave, whereas the direct EWP serves as a reference wave. Their coherent superposition in the far field gives rise to characteristic interference structures in the PMD, which encode information about the electron dynamics and the target structure~\cite{Remetter2006_NatPhys}. Among these structures, FSH interference produces characteristic \textit{spider-leg-like} patterns through the interference of signal and reference EWPs ionized within the same quarter-cycle (QC) of the laser field. These structures provide information about the continuum EWP and the ionization dynamics~\cite{Huismans2011_Science,Meckel2014_NatPhys,Hickstein2012_PRL}. In contrast, BSH interference gives rise to characteristic \textit{fish-bone-like} structures through the interference of EWPs originating from different QCs, and are particularly sensitive to the molecular structure~\cite{Bian2012_PRL}. In addition to these holographic contributions, non-holographic time-double-slit (TDS) interference can arise from the coherent superposition of two direct EWPs ionized during adjacent QCs and reaching the detector without rescattering~\cite{Gopal2009_PRL}.

For laser pulses extending over more than one optical cycle, EWPs can be ionized during multiple QCs, resulting in the simultaneous contribution of several holographic and non-holographic pathways to the PMD. The resulting overlap of different interference mechanisms can make it difficult to unambiguously identify the underlying electron trajectories and isolate structure-sensitive holographic features. Sub-cycle laser pulses provide a means of reducing this pathway complexity by temporally confining the ionization process and limiting the number of relevant intracycle electron trajectories. They therefore offer a promising route toward resolving sub-cycle electron dynamics and isolating structure-relevant holographic interference patterns~\cite{Du2016_OptLett,Murakami2020_PRA,Taoutioui2022_SciRep,Khurelbaatar2024_LightSci}. Recent advances in ultrafast laser technology have enabled the generation of pulses with full-width at half-maximum (FWHM) durations shorter than one optical cycle~\cite{Nishimiya2025_AdvPhysX,Xu2024_NatPhoton,Nishimiya2024_OptLett,Rossi2020,Nie2018_NatPhoton,Liang2017,Chu_2016,Balciunas2015_NatCommun,Voronin2013_OptLett,Krauss2010_NatPhoton}. Such sub-cycle pulses have also attracted considerable interest for the generation and control of attosecond light fields~\cite{Holkundkar2023_PLA,Rajpoot2025_PRR,Nishimiya2026_SciAdv,Rajpoot2020_JPhysB}. In the context of SPH, the enhanced temporal localization provided by sub-cycle pulses can help distinguish contributions from different ionization and rescattering pathways and thereby facilitate a clearer interpretation of the resulting photoelectron holograms~\cite{Yuan2021_PRA,Murakami2020_PRA,Khurelbaatar2024_LightSci,Rajpoot2026_PRR}.

Polar molecules provide an especially interesting platform for SPH because their asymmetric charge distributions and permanent dipole moments can influence both the initial ionization and the subsequent propagation and rescattering of the EWP~\cite{Bian2014_PRA,Wang2020_PRA,Habibovic2021_PRA}. The interplay between the molecular dipole, the ionic Coulomb potential, and the molecular orientation can therefore modify the resulting photoelectron interference patterns. Recent SPH measurements of HCl have resolved subcycle dynamics associated with ionization from the HOMO and HOMO-1 orbitals and revealed orientation-dependent asymmetries in the PMDs~\cite{Haertelt2026_PRL}. These findings motivate a systematic investigation of the influence of molecular orientation on different holographic interference pathways in polar molecules. Particular attention is given to BSH interference, which is sensitive to molecular structure but can be difficult to isolate from competing interference contributions under few-cycle or longer-pulse driving conditions. Sub-cycle laser pulses may therefore provide a means of resolving BSH features more clearly and investigating their sensitivity to molecular structure.

In this work, we investigate photoelectron holographic interference in the polar CO molecule driven by a half-cycle FWHM laser pulse using two-dimensional time-dependent \Sch equation (2D-TDSE) simulations combined with a classical-trajectory-based analysis. We first examine the dependence of the FSH interference on molecular orientation and relate the observed fringe shifts to the orientation-dependent transverse displacement of the ionizing EWP. We then identify the curved interference structures in the negative parallel-momentum region as BSH interference and analyze the corresponding intracycle electron trajectories. Particular attention is given to the dependence of the BSH interference on molecular orientation and its robustness against orientation averaging. We further investigate the influence of the CEP on the BSH interference, including its persistence after CEP averaging. Finally, we examine the dependence of the BSH interference pattern on the internuclear separation to assess its sensitivity to molecular geometry. These analyses establish the relationship between molecular orientation, electron-trajectory pathways, and the resulting holographic interference structures, and highlight the potential of sub-cycle BSH as a probe of molecular structure under conditions of incomplete molecular orientation and finite CEP variations.

The paper is organized as follows. Section~\ref{sec2} describes the theoretical methods and the sub-cycle laser-field profile employed in the simulations. Section~\ref{sec3} presents and analyzes the resulting photoelectron holographic interference structures, including their dependence on molecular orientation, CEP, and internuclear separation. Section~\ref{sec4} summarizes the main findings and provides concluding remarks. Atomic units (a.u.) are used throughout unless otherwise stated.
  
\section{Theoretical Methods}
\label{sec2} 
In this section, we describe the theoretical framework employed for the calculation of PMDs. We first introduce the molecular potential and numerical procedure used to solve the TDSE, followed by a discussion on the sub-cycle field profile used in this work.

\subsection{Simulation details for TDSE}
\label{sec2_tdse}
\begin{figure}[t]
\centering\includegraphics[width=0.6\columnwidth]{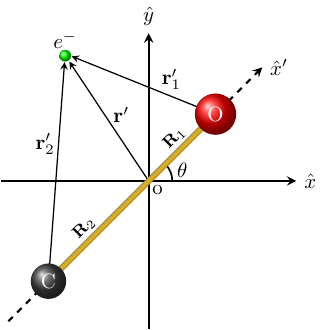}
\caption{Schematic illustration of the coordinate system and molecular geometry of the CO molecule used in the simulation. The molecular axis $x'$ is oriented at an angle $\theta$ with respect to the laser-field polarization axis $x$. The vectors $\mb{R}_1$ and $\mb{R}_2$ denote the positions of the  O and C nuclei, respectively, measured from the molecular center of mass, which coincides with the origin $o$ of the coordinate system. The vectors $\mb{r}'$, $\mb{r}_1'$, and $\mb{r}_2'$ represent the positions of the electron $e^-$ relative to the origin $o$, the O nucleus, and the C nucleus, respectively. }
\label{fig_mol_frame}
\end{figure}

We investigate the PMD of a carbon monoxide (CO) molecule in two spatial dimensions. The effective molecular potential in the single-active-electron (SAE) approximation is given by~\cite{Nalda2004_PRA, Chen2014_PRL, Wang2017_PRA},
\be
V(\mb{r}') = -  \mathlarger{\sum_{j=1,2}}\frac{Z_j^o + (Z_j^i - Z_j^o) \exp(-\rho \mb{r}_j'^2)}{\sqrt{\mb{r}_j'^2 + a_0}},
\ee
where $\mb{r}' = (x',y')$ denotes the molecular-frame coordinates. The coordinate system and molecular geometry used in the simulation are illustrated in Fig.~\ref{fig_mol_frame}. The molecular and laboratory frames are related through the rotation
\be 
x = x'\cos\theta + y'\sin\theta,\qquad y = -x'\sin\theta + y'\cos\theta,
\ee
where $\theta$ is the angle between the molecular axis and the polarization direction of the laser field. The index $j = 1,2$ labels the O and C nuclei, respectively, with effective nuclear charges characterized by the inner and outer charge parameters $Z_j^i$ and $Z_j^o$. The quantity $\mb{r}_j'^2 = |\mb{r}' - \mb{R}_j|^2$, where $\mb{R}_j = (\pm R_j,0)$ denotes the position of the corresponding nucleus, with $j=1$ ($2$) associated with the ``$+$'' (``$-$'') sign.  The distances of the nuclei from the center of mass are given by 
\be 
R_1 = \frac{Z_2 R}{Z_1 + Z_2},\qquad R_2 = \frac{Z_1 R}{Z_1 + Z_2},
\ee
where $R$ is the internuclear separation. The parameters $\rho$ and $a_0$ represent the screening and softening parameters, respectively. For the CO molecule, we use the parameters $Z_1^i = 6$, $Z_2^i = 4$, $Z_1^o = 0.6$, $Z_2^o = 0.4$, $R = 2.135$ a.u., $\rho = 1.746$, and $a_0 = 0.69$~\cite{Nalda2004_PRA, Chen2014_PRL, Wang2017_PRA}. These parameters yield an ionization potential of $I_p = 0.52$ a.u. for the highest occupied molecular orbital (HOMO), which has $\sigma$-type orbital symmetry.

The PMDs are obtained by numerically solving the 2D-TDSE within the SAE and dipole approximations. In the length gauge, the TDSE is written as
\be
	i \frac{\partial}{\partial t} \psi(\mb{r},t) = \left[- \frac{1}{2}\nabla^2 + V(\mb{r}) + \mb{r}\cdot\mb{E}(t) \right] \psi(\mb{r},t),
\label{eq_tdse}	
\ee
where $\psi(\mb{r},t)$ is the time-dependent electronic wave function and $\mb{E}(t)$ is the electric field of the laser pulse. The initial bound state is obtained by imaginary-time propagation of the field-free Hamiltonian. The subsequent real-time evolution of the wave function is performed using the split-operator method~\cite{Feit1982_JCP,Rajpoot2020_JPhysB}. 

To reduce the computational domain required to describe the propagation of the ionized wave packet, we employ the wave-function splitting method~\cite{Chelkowski1998_PRA}. In this approach, the wave function is separated into inner and outer components according to
\be \psi(\mb{r},t) = \psi_\text{in}(\mb{r},t) + \psi_\text{out}(\mb{r},t),\ee
\be \psi_\text{in}(\mb{r},t) = \left[1-M(r)\right] \psi(\mb{r},t),\label{eq_mask_in}\ee
\be \psi_\text{out}(\mb{r},t) = M(r) \psi(\mb{r},t),\label{eq_mask_out}\ee  
where $M(r)$ is a radial mask function defined as
\be 
M(r) = \frac{1}{1 + \exp[-(r-r_c)/\eta]}.
\ee
Here, $r = \sqrt{x^2+y^2}$, $\eta = 10$ determines the smoothness of the mask, while $r_c = 130$ a.u. defines the onset of the absorbing region. At this distance, the Coulomb force is negligible compared with the force exerted by the laser field. 

The inner component $\psi_\text{in}(\mb{r},t)$ is numerically propagated under the full Hamiltonian. At intervals of $5$ a.u., the portion of the wave function entering the outer region is extracted using Eq.~\ref{eq_mask_out}. The extracted wave packet is subsequently transformed from length gauge to the velocity gauge according to
\be 
\psi_\text{out}^{VG}(\mb{r},t) = \exp\left[-i \mb{A}(t) \cdot \mb{r} \right] \psi_\text{out}(\mb{r},t),
\ee
where $\mb{A}(t) = -\int_0^t \mb{E}(t') dt'$ is the laser vector potential. The velocity-gauge wave packet is then Fourier transformed to momentum space to obtain $\psi_\text{out}^{VG}(\mb{p},t)$. Each momentum-space wave packet is analytically propagated to a final time $t_f$ according to the corresponding Volkov, i.e., free electron in the laser field, evolution. The final PMD amplitude is obtained by coherently accumulating the contributions from all extracted wave packets propagated to time $t_f$. The PMD is initially calculated in the molecular frame and subsequently transformed to the laboratory frame using 
\be 
p_x = p_x'\cos\theta + p_y'\sin\theta,\qquad p_y = -p_x'\sin\theta + p_y'\cos\theta.
\ee
where $(p_x',p_y')$ and $(p_x,p_y)$ denote the momentum components in the molecular and laboratory frames, respectively.
 
For all simulations, a Cartesian spatial grid extending from $-200$ to $200$ a.u. is used in both the $x$ and $y$ directions, with spatial steps $\Delta x = \Delta y = 0.2$ a.u. and a time step of $\Delta t = 0.05$ a.u. The numerical convergence of the calculations has been verified by varying the grid boundary, $r_c$, $\Delta x$, $\Delta y$, and the time step, $\Delta t$. After the laser pulse is switched off, the wave function is propagated for an additional four optical cycles, allowing the ionized wave packet to sufficiently separate from the ionic core before the final PMD is evaluated.

\subsection{Sub-cycle field profile}
\label{sec2_laser}
We describe the linearly polarized laser pulse using the sub-cycle pulsed beam (SCPB) model, which allows the pulse duration to be varied continuously, including the sub-cycle regime~\cite{LinSubcycle-2006}. The SCPB are derived from the oscillating dipole model using the complex-source point method \cite{PhysRevE.67.016503,Heyman-89}. In the dipole approximation and plane-wave limit, the electric field of the linearly polarized pulse is expressed as
\be
\mb{E}(t) = \text{Re}\left\{ \frac{A_d(t')}{|A_d(0)|} f(t')\ E_0\exp(i\omega_0t' + i\phi_0) \right\} \hat{\mb{e}}_x,
\label{laser}
\ee
where $E_0$ is the peak field amplitude, $\omega_0$ is the carrier frequency, and $\phi_0$ is the carrier-envelope phase (CEP). The retarded time is defined as $t' = t - t_0 - z/c$, where $t_0$ denotes the pulse center. Since the simulations are performed in the dipole approximation, we set $z = 0$. The complex, time-dependent correction factor $A_d(t')$ is given by~\cite{Cai2016_OptCommun}:
\be
A_d(t') = 1 - i \dot{f}(t')/[\omega_0 f(t')],
\ee 
where $\dot{f}(t')$ denotes the first derivative of the envelope function with respect to $t'$. In the present work, an analytical Gaussian envelope is used to describe the sub-cycle pulse profile~\cite{Cai2016_OptCommun},
\be 
f(t') = e^{-(t'/T)^2} f_1 / |f_1|,
\label{laser_envlp}
\ee
where 
\be 
f_1 = 1 + \frac{2i}{\sqrt{\pi} T} \int_0^{t'} e^{(t''/T)^2} dt'',
\ee
and $|f_1|$ provides the corresponding normalization. The parameter $T = \tau/\sqrt{2\text{ln}2}$, where $\tau$ is the FWHM of the intensity envelope function $|f(t')|^2$. Throughout this work, we consider a linearly polarized pulse with wavelength of $\lambda_0 = 800$ nm, corresponding to $\omega_0 = 0.057$ a.u., and a peak intensity of $I_0 = 2\times 10^{14}$ W$/$cm$^{2}$. For these parameters, ionization of the CO molecule occurs predominantly in the tunneling regime, as characterized by the Keldysh parameter $\gamma = \omega_0 \sqrt{2I_p}/E_0 < 1$~\cite{Keldysh1965_JETP}. Unless otherwise specified, the CEP is set to $\phi_0 = 0^\circ$, the intensity-envelope FWHM is $\tau = 0.5T_0$, and the pulse is defined over a total duration of $3T_0$. The wave function is subsequently propagated for an additional $4T_0$ after the pulse has ended, resulting in the final simulation time of $t_f = 7T_0$. Here, $T_0 = 2\pi/\omega_0$ denotes the optical period of the laser field.

\section{Results and Discussion}
\label{sec3}
In this section, we first discuss the salient features of the PMDs of the CO molecule generated by a half-cycle laser pulse. In particular, we identify the FSH and BSH interference structures and distinguish them from the direct-electron contribution. We then examine the evolution of these holographic interference features with molecular orientation relative to the laser-field polarization direction. Subsequently, we investigate the robustness of the BSH interference against molecular-orientation averaging and variations in the sub-cycle pulse CEP, including CEP averaging. Finally, we examine the sensitivity of the BSH interference pattern to the internuclear separation.
\begin{figure}[t]
\centering\includegraphics[width=0.85\columnwidth]{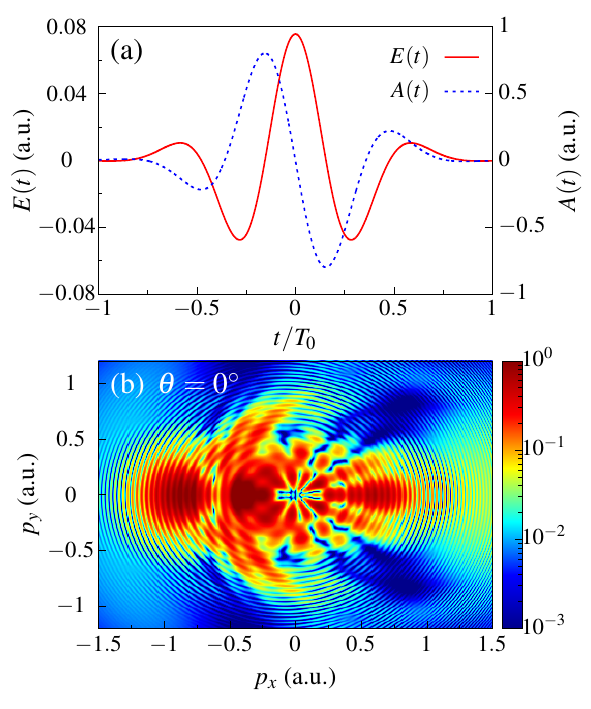}
\caption{ (a) Temporal profiles of the laser electric field $\mb{E}(t)$ (red solid curve, left axis) and vector potential $\mb{A}(t)$ (blue dashed curve, right axis) for a Gaussian-enveloped pulse with FWHM $\tau = 0.5T_0$. (b) Corresponding TDSE-simulated PMD of CO molecule at orientation angle $\theta = 0^\circ$. } 
\label{fig2}
\end{figure}

\subsection{PMD characteristics for half-cycle pulse}
\label{subsec1}
Figure~\ref{fig2}(a) shows the temporal profiles of the electric field $\mb{E}(t)$ (red solid curve) and vector potential $\mb{A}(t)$ (blue dashed curve) for the half-cycle Gaussian laser pulse considered in this work. The corresponding TDSE-simulated PMD for a CO molecule aligned parallel to the laser polarization direction, $\theta = 0^\circ$, is shown in Fig.~\ref{fig2}(b). In this configuration, the C and O nuclei are located on the negative and positive $x$ sides, respectively, and $p_{x}$ denotes the momentum component along the laser polarization direction.

The PMD exhibits characteristic holographic interference structures associated with linearly polarized strong laser fields~\cite{Bian2011_PRA, Marchenko2011_PRA, Rajpoot2026_PRR}. In the $p_x > 0$ region, a spider-leg-like interference pattern is observed, which is attributed to FSH interference. In contrast, the $p_x < 0$ region exhibits a set of curved fringes associated with BSH interference. Remarkably, the BSH fringes are clearly resolved in the present half-cycle pulse. In longer pulses, such structures can become obscured by the superposition of contributions from intercycle and non-holographic interference. Because backward-scattered electrons approach the ionic core with relatively small impact parameter, their interference patterns can provide enhanced sensitivity to the molecular potential and, consequently, valuable structural information about the target. 

In addition to the holographic structures, a pronounced PMD feature is observed in the momentum range $-1.1 < p_x < -0.7$ a.u. This contribution is predominantly associated with direct electrons ionized near the positive maximum of the vector potential, which occurs at $t = -0.15T_0$, where $A(t = -0.15T_0) \approx 0.8$ a.u. For an electron ionized at time $t_i$ with negligible initial velocity and subsequently driven solely by the laser field, its final drift momentum is given by $\mb{p}_f = -\mb{A}(t_i)$. Consequently, electrons ionized near this maximum of the vector potential acquire a negative final momentum of approximately $p_{f,x} \approx -A(t = -0.15T_0) \approx -0.8$ a.u., without undergoing rescattering from the molecular core. The appreciable intensity of this direct-electron contribution can be further understood from the asymmetric ionization of the oriented CO molecule. For $\theta=0^\circ$, the C nucleus is located on the negative-$x$ side, and the positive electric field drives the electron toward the negative-$x$ direction, i.e., ionization occurs from the C side of the molecule. Moreover, the molecular dipole is anti-parallel to the applied field in this half-cycle, leading to field-induced dressing of the ground state and enhanced ionization from the C side~\cite{Holmegaard2010_NatPhys, Dimitrovski2010_PRA, Wang2020_PRA, Wu2012_PRL}.

\subsection{Effect of molecular orientation on FSH interference}
\label{subsec2}
\begin{figure}[t]
\centering\includegraphics[width=1.0\columnwidth]{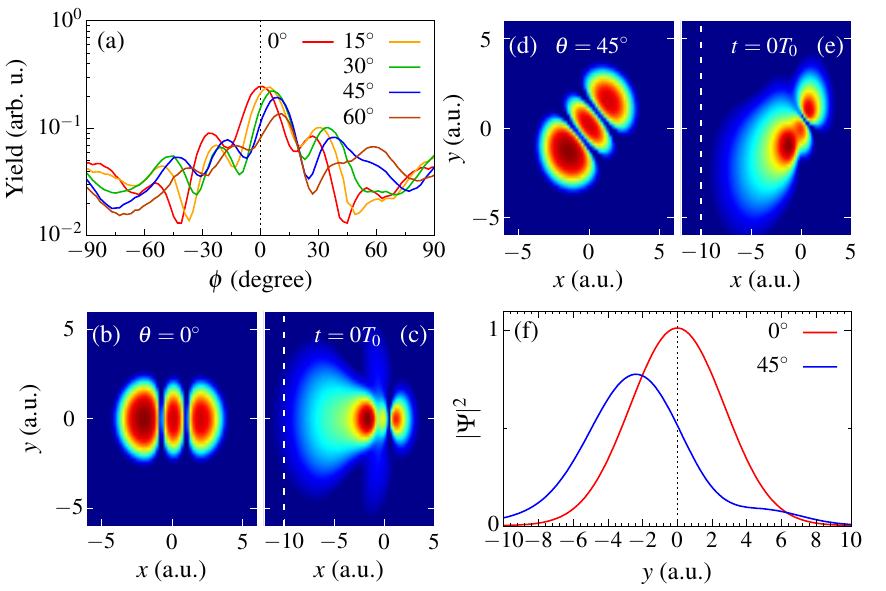}
\caption{ (a) Photoelectron momentum angular distributions obtained from TDSE-simulated PMDs of CO for different molecular orientation angles $\theta$. (b) and (c) Electron density distributions of CO in coordinate space for $\theta = 0^\circ$ at the beginning of the laser pulse ($t=-1T_0$) and at the peak of the pulse ($t=0T_0$), respectively. (d) and (e) Same as (b) and (c), respectively,  but for $\theta = 45^\circ$. (f) Line profiles of the electron density distributions for $\theta = 0^\circ$ and $45^\circ$, extracted from panels (c) and (e), respectively, along the dashed line at $x = -10$ a.u. indicated in panels (c) and (e). The line profiles are normalized such that the maximum value for $\theta=0^\circ$ is unity. } 
\label{fig3}
\end{figure}
Next, we investigate the influence of the molecular orientation angle $\theta$ on the FSH interference fringes. Figure~\ref{fig3}(a) shows the photoelectron momentum angular distributions (PMADs) of CO for different molecular orientations. The PMADs are obtained by radially integrating the TDSE-simulated PMDs over the momentum range $0<p<1.5$ a.u. and are presented as a function of the polar angle $\phi$ [$=\tan^{-1}(p_y/p_x)$] over the range $-90^\circ \leq \phi \leq 90^\circ$. For a molecular orientation parallel to the laser polarization axis, $\theta = 0^\circ$, the PMAD is symmetric about $\phi = 0^\circ$, with its maximum intensity located at $\phi = 0^\circ$. As $\theta$ increases, the PMAD becomes increasingly asymmetric, accompanied by a shift of the maximum intensity region toward larger $\phi$. This angular shift of the FSH interference pattern can be associated with the non-zero transverse displacement and the corresponding phase structure of the ionizing EWP, which depend on the molecular orientation relative to the laser polarization~\cite{Tan2023_PRA, Meckel2014_NatPhys, Liu2016_PRL}.

To examine the origin of this orientation-dependent shift, we analyze the electron density in coordinate space. Figures~\ref{fig3}(b) and \ref{fig3}(c) show the electron density distributions for $\theta = 0^\circ$ at the beginning of the laser pulse, $t = -1T_0$, and at the peak of the pulse, $t = 0T_0$ [see Fig.~\ref{fig2}(a)], respectively. The corresponding contributions for $\theta = 45^\circ$ are shown in Figs.~\ref{fig3}(d) and \ref{fig3}(e). Since the FSH interference is primarily formed by EWPs ionized within the laser quarter-cycle following the peak of the driving field, the electron density near the ionization region at the peak of the pulse provides an indication of the initial spatial distribution of the ionizing EWP. For a more quantitative comparison, the corresponding line profiles along $x = -10$ a.u. are shown in Fig.~\ref{fig3}(f). For $\theta = 0^\circ$, the electron density is symmetric with respect to the $x$ axis and reaches its maximum at $y = 0$ a.u. In contrast, for $\theta = 45^\circ$, the maximum of the electron density is displaced to approximately $y = -2.5$ a.u. This non-zero transverse displacement reflects the anisotropic spatial distribution of the molecular orbital when the molecular axis is oriented with respect to the laser polarization axis~\cite{Tan2023_PRA}. Such an orientation-dependent initial transverse displacement is expected to influence the emission angle of the EWP and, consequently, the position of the FSH interference fringes in the PMAD.
 
\begin{figure}[t]
\centering\includegraphics[width=1.0\columnwidth]{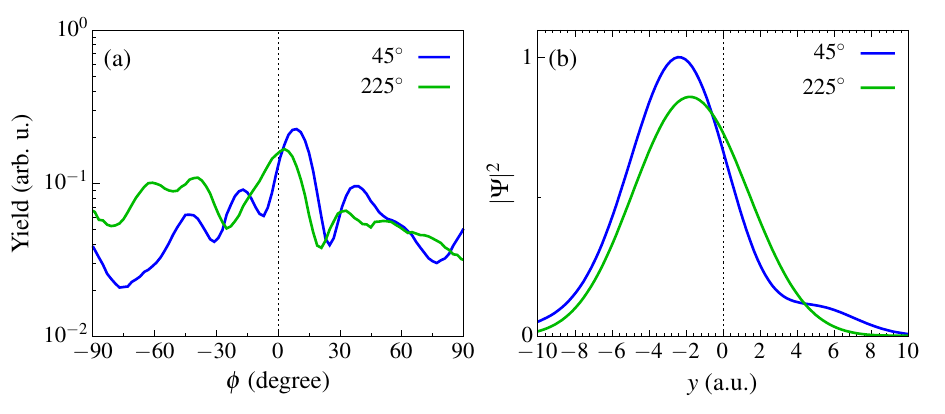}
\caption{ (a) Photoelectron momentum angular distributions obtained from the TDSE-simulated PMDs of CO for molecular orientations $\theta = 45^\circ$ (blue curve) and $225^\circ$ (green curve). (b) The corresponding line profiles of the electron density distributions, extracted at $x = -10$ a.u., for the two molecular orientations. The line profiles are normalized such that the maximum value for $\theta=45^\circ$ is unity. } 
\label{fig4}
\end{figure}
Furthermore, we examine the role of the heteronuclear nature of CO by comparing the PMADs for molecular orientations $\theta = 45^\circ$ and $225^\circ$, as shown in Fig.~\ref{fig4}(a). These two orientations correspond to ionization predominantly from the opposite sides of the molecule: for $\theta = 45^\circ$, the C side oriented toward the direction of electron emission, whereas for $\theta = 225^\circ$ the O side is oriented toward this direction. The PMADs exhibit a different degree of angular shift of the FSH interference fringes for these two orientations. In particular, the shift is more pronounced for $\theta=45^\circ$, where ionization occurs predominantly from the C side. The corresponding line profiles of electron density distributions are shown in Fig.~\ref{fig4}(b). The electron density associated with $\theta=45^\circ$ exhibits a larger transverse displacement in the ionization region than that for $\theta=225^\circ$. This difference in the initial transverse position of the ionizing EWP is consistent with the different shifts of the FSH interference fringes observed in Fig.~\ref{fig4}(a), highlighting the influence of the asymmetric electronic structure of the heteronuclear CO molecule on the holographic interference pattern.

\subsection{BSH interference from CO molecule}
\label{subsec3}
\begin{figure}[t]
\centering\includegraphics[width=1.0\columnwidth]{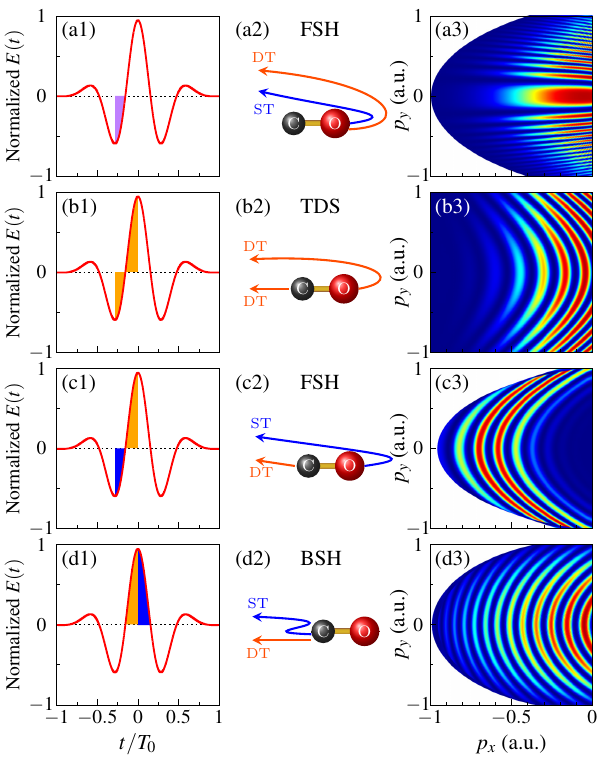}
\caption{ Schematic illustration of the different intracycle interference pathways and the corresponding interference structures calculated using the classical-trajectory based model. Left column: laser electric field profiles, with the same parameters as those used in Fig.~\ref{fig2}(a). The shaded quarter-cycles indicate the ionization times associated with the interfering trajectories: orange and blue denote the direct (DT) and scattering (ST) trajectories, respectively, while purple denotes the common ionization quarter-cycle for the two trajectories in the FSH scenario. Middle column: schematic representations of the corresponding interfering electron trajectories originating from different sides of the CO molecule. Right column: calculated holographic and non-holographic interference structures associated with the different ionization pathways. Panels (a), (b), (c), and (d) correspond to the FSH, TDS, FSH, and BSH interference scenarios, respectively.} 
\label{fig5}
\end{figure}

We next examine the BSH interference fringes observed in the negative parallel-momentum region ($p_x<0$) of the PMD shown in Fig.~\ref{fig2}(b). To identify the origin of these fringes, we consider the possible intracycle interference pathways that can contribute to the negative-$p_x$ region for the half-cycle laser pulse. Figure~\ref{fig5} illustrates the relevant interference scenarios using a classical-trajectory based model~\cite{Rajpoot2026_PRR}, which  provides a qualitatively description of the interference structures observed in the TDSE calculations. The corresponding laser-field profiles, interfering electron trajectories, and calculated interference patterns are shown in the left, middle, and right columns, respectively.

Figure~\ref{fig5}(a) illustrates an FSH interference scenario in which the interfering EWPs are ionized within the same QC, indicated by the purple-shaded region in panel (a1). The corresponding interference pattern in panel (a3) exhibits the characteristic spider-leg-like structure associated with FSH. Figure~\ref{fig5}(b) represents a non-holographic time double-slit (TDS) interference scenario~\cite{Bian2011_PRA, Gopal2009_PRL, Rajpoot2026_PRR}. In this case, two direct trajectories (DTs) are ionized during different QCs and from opposite sides of the CO molecule. Neither trajectory undergoes rescattering from the molecular core, but the two trajectories can arrive at the detector with the same final momentum. Their coherent superposition gives rise to the TDS interference pattern shown in Fig.~\ref{fig5}(b3). 

Neither the FSH nor the TDS interference patterns shown in Figs.~\ref{fig5}(a3) and \ref{fig5}(b3) provide the dominant contribution to the TDSE-simulated PMD in Fig.~\ref{fig2}(b). In both cases, atleast one of the interfering EWPs is ionized from the O side of the molecule. For the present molecular orientation ($\theta=0^\circ$), the molecular dipole moment is directed along the negative $x$ direction, from O to C. During the relevant negative-QC of the laser field, the molecular dipole is parallel to the applied electric field. This field configuration leads to field-induced dressing of the HOMO and increases the effective ionization energy, thereby suppressing ionization from the O side~\cite{Holmegaard2010_NatPhys, Dimitrovski2010_PRA, Wang2020_PRA, Wu2012_PRL}. Moreover, the electric-field amplitude during this QC is relatively weak compared with that during the subsequent QCs near the peak of the pulse. Consequently, the O-side ionization pathways required for the FSH and TDS scenarios in Figs~\ref{fig5}(a) and \ref{fig5}(b) have relatively small contribution, making the corresponding interference structures weak or absent in the PMD.

Figure~\ref{fig5}(c) shows another FSH interference scenario. Here, the orange- and blue-shaded QCs in panel (c1) indicate the ionization times of the direct (DT) and scattering (ST) trajectories, respectively, whose interference produces the patterns shown in panel (c3). The resulting structures consists of right-curved fringes that resemble those observed in the negative-$p_x$ region of the TDSE PMD in Fig.~\ref{fig2}(b). However, the scattering trajectory in this scenario originates from the O side of the molecule and is ionized during a relatively weak QC of the laser field. Thus, similar to the cases discussed above, this pathway has a comparatively small ionization probability and is not expected to make a significant contribution to the observed TDSE PMD.

Finally, we consider the BSH interference scenario shown in Fig.~\ref{fig5}(d). In this case, the two interfering EWPs originate predominantly from the C side of the molecule during the two strong QCs located on either side of the peak of the laser pulse. During these QCs, the applied electric field is anti-parallel to the molecular dipole moment. The resulting field-induced dressing lowers the effective ionization energy and enhances ionization from the C side~\cite{Wang2020_PRA,Wu2012_PRL}. Thus, in contrast to the interference pathways involving O-side ionization, both EWPs contributing to the BSH  interference are associated with relatively strong ionization channels. The corresponding calculated interference pattern in Fig.~\ref{fig5}(d3) exhibits right-curved fringes that closely resemble those observed in the negative-$p_x$ region of the TDSE PMD in Fig.~\ref{fig2}(b). These observations support the identification of the right-curved interference fringes in the negative-$p_x$ region as BSH interference arising from the coherent superposition of direct and backward-scattered electron wave packets.

\subsection{Effect of molecular orientation on BSH interference}
\label{subsec4}
\begin{figure}[t]
\centering\includegraphics[width=1.0\columnwidth]{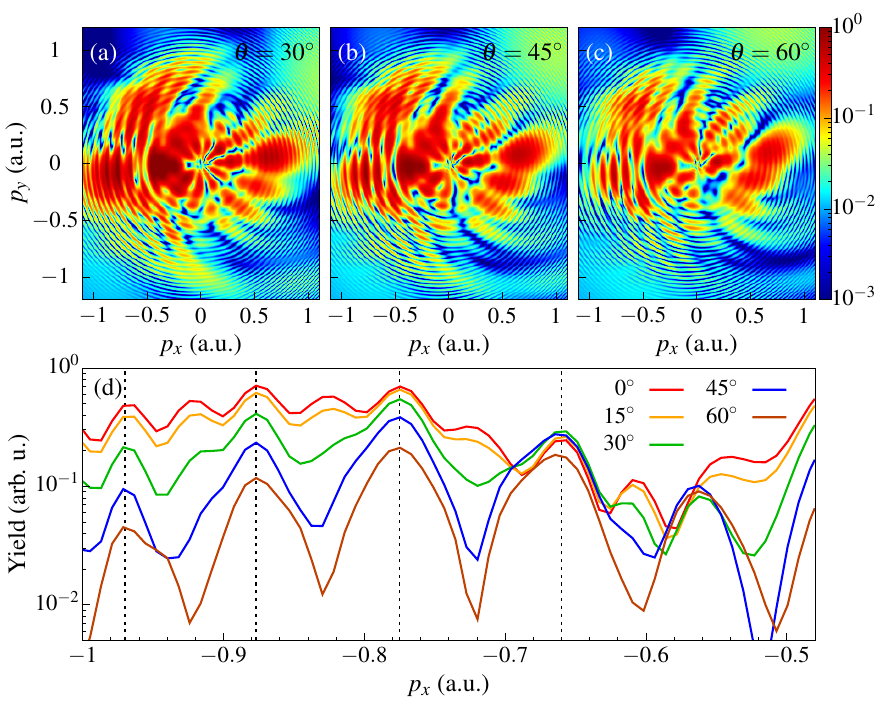}
\caption{ TDSE-simulated PMDs of CO for different molecular orientation angles $\theta:$ (a) $30^\circ$, (b) $45^\circ$, and (c) $60^\circ$, calculated using a half-cycle driving pulse with the same laser parameters as those used in Fig.~\ref{fig2}(a). (d) Line profiles of the BSH interference structure extracted along $p_y=0$ a.u. from the corresponding PMDs for different orientation angles. The vertical dotted lines indicate the positions of the BSH interference peaks.} 
\label{fig6}
\end{figure}
Having established the origin of the BSH interference structures in the PMDs, we next investigate their dependence on the molecular orientation angle $\theta$. Figures~\ref{fig6}(a)--(c) show the TDSE-simulated PMDs of CO for $\theta=30^\circ$, $45^\circ$, and $60^\circ$, respectively. The laser parameters are the same as those used in Fig.~\ref{fig2}(a). The visibility of the BSH interference structures increases systematically with increasing $\theta$. This behavior is consistent with an increasing contribution from backward-scattered electrons as the molecular orientation changes, leading to a stronger BSH contribution relative to the direct-electron background. Consequently, the direct-electron distribution in the momentum range $-1.1<p_x<-0.7$ a.u., which is prominent for $\theta=0^\circ$, becomes increasingly dominated by the BSH interference fringes at larger orientation angles.

A more quantitative comparison is provided by the line profiles shown in Fig.~\ref{fig6}(d), which are extracted from the TDSE-simulated PMDs along $p_y=0$ a.u. The increasing contrast of the BSH peaks with $\theta$ reflects the enhanced relative contribution of the backward-scattered EWP. For $\theta=0^\circ$ and $15^\circ$, additional structures are observed in the line profiles because the BSH contribution overlaps with the direct-electron distribution. This overlap becomes progressively less pronounced as $\theta$ increases, allowing the BSH interference fringes to become more clearly resolved.

Interestingly, in contrast to the FSH interference fringes, whose peak positions shift with molecular orientation, the positions of the BSH interference peaks remain nearly unchanged over the range of orientation angles considered, as indicated by the vertical dotted lines in Fig.~\ref{fig6}(d). This difference can be understood from the distinct geometries of the two interference structures. The BSH fringes form arc-like structures in the PMD. Upon rotating the molecular axis, these structures can change their overall distribution around the center of the PMD while their intersections with the $p_y=0$ axis, and hence the peak positions extracted along this axis, remain nearly unchanged. In contrast, the spider-leg-like FSH structures are sensitive to the initial transverse displacement of the ionizing wave packet. As discussed in relation to Fig.~\ref{fig3}, this displacement changes with molecular orientation, resulting in a corresponding shift of the FSH fringe positions.

The relative insensitivity of the BSH fringe positions to molecular orientation is particularly relevant for experiments involving incompletely aligned molecular ensembles. The persistence of well-defined BSH features over a range of molecular orientations suggests that these structures may provide access to molecular-structure information even when precise molecular alignment is not achieved.

\begin{figure}[t]
\centering\includegraphics[width=1\columnwidth]{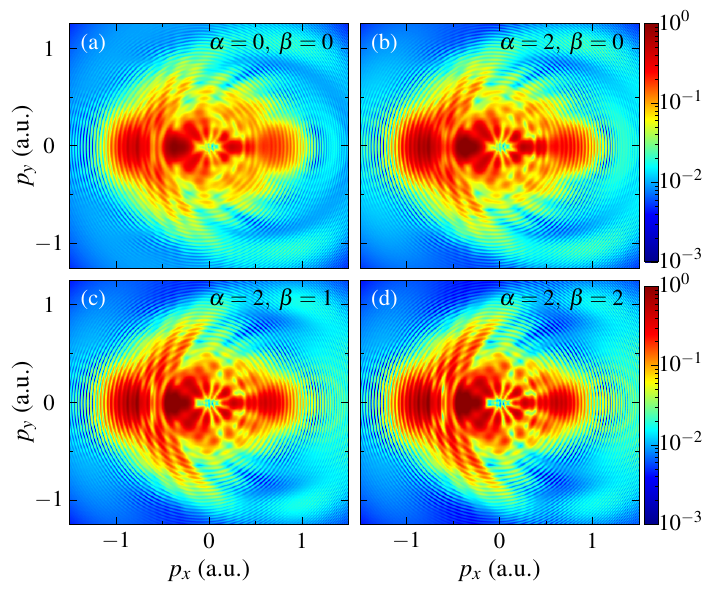}
\caption{Angle-averaged PMDs of CO for different molecular-orientation distributions characterized by the parameters $\alpha$ and $\beta$ in Eq.~\ref{eq_weight_function}. (a) Uniform angular distribution ($\alpha = \beta = 0$); (b) preferential alignment without net orientation ($\alpha=2,\beta=0$); (c) preferential alignment and orientation ($\alpha=2,\beta=1$); and (d) stronger orientation bias ($\alpha=2,\beta=2$). The corresponding degrees of alignment, $\langle\cos^2\theta\rangle$, and orientation, $\langle\cos\theta\rangle$, are listed in Table~\ref{tab_costh}. }
\label{fig7}
\end{figure}
To further examine the robustness of the BSH interference against molecular-orientation averaging, we calculate the angle-averaged PMD of CO for molecular ensembles whose molecular axes lie in the $xy$-plane. The molecular orientation is characterized by the angle $\theta \in [0,2\pi]$ measured from the laser polarization direction. To account for a nonuniform population of molecular orientations, we introduce the angular distribution
\be 
W(\theta) = \frac{1}{N} \exp\left[\alpha\cos^2\theta + \beta\cos\theta\right],
\label{eq_weight_function}
\ee
where $N = \int_0^{2\pi} \exp[\alpha\cos^2\theta + \beta\cos\theta] d\theta$ ensures the normalization $\int_0^{2\pi} W(\theta) d\theta = 1$. The parameters $\alpha (\geq 0)$ and $\beta$ control the degree of molecular alignment and the head-to-tail orientation bias. Specifically, a positive $\alpha$ favors molecular axes parallel or anti-parallel to the laser polarization, whereas a nonzero $\beta$ introduces a preference for one of these two orientations over the other. The degree of alignment is characterized by the expectation value $\langle \cos^2 \theta \rangle = \int_0^{2\pi} W(\theta) \cos^2\theta d\theta$, which approaches unity for perfect alignment along the polarization axis and equals $1/2$ for a uniform angular distribution in the $xy$-plane. The degree of  orientation is quantified by $\langle \cos\theta \rangle = \int_0^{2\pi} W(\theta) \cos\theta d\theta$. This quantity ranges from $-1$ to $1$, with $\langle\cos\theta\rangle = \pm 1$ corresponding to  perfect orientation along either direction of the polarization axis, while $\langle\cos\theta\rangle = 0$ indicates no net head-to-tail orientation. Note that the latter condition does not necessarily imply a uniform angular distribution, as an aligned ensemble may contain equal populations of parallel and anti-parallel molecular orientations.

The angle-averaged PMD is obtained through a weighted averaging of the orientation-resolved distributions according to
\be
P_\text{avg}(p_x,p_y) =  \int_{0}^{2\pi} W(\theta) P_{\theta}(p_x,p_y)d\theta,
\label{eq_angle_average}
\ee
where $P_{\theta}(p_x,p_y)$ denotes the PMD calculated for a fixed molecular orientation $\theta$. The integration over the full $2\pi$ range accounts for the distinct head-to-tail orientations of the heteronuclear CO molecule within the assumed planar ensemble. This averaging procedure allows us to examine how preferential molecular alignment and orientation affect the visibility and persistence of photoelectron holographic interference structures in the ensemble-averaged PMD.

\begin{table}[t]
\setlength{\tabcolsep}{2pt}
\renewcommand{\arraystretch}{1.3}
\centering
\caption{\label{tab_costh} Parameters $\alpha$ and $\beta$ defining the molecular-orientation distributions used in Fig.~\ref{fig7}, together with the corresponding degrees of alignment $\langle\cos^2\theta\rangle$ and orientation $\langle\cos\theta\rangle$. }
\begin{ruledtabular}
\begin{tabular}{cccc}
$\alpha$ & $\beta$ & $\langle\cos^2\theta\rangle$ & $\langle\cos\theta\rangle$\\
\colrule
0 & 0 & 0.50 & 0.00  \\ 
2 & 0 & 0.72 & 0.00  \\ 
2 & 1 & 0.76 & 0.60  \\
2 & 2 & 0.81 & 0.83 \\
\end{tabular}
\end{ruledtabular}
\end{table}

Figure~\ref{fig7} presents the resulting angle-averaged PMDs for different angular distributions. The PMD in Fig.~\ref{fig7}(a) corresponds to $\alpha=\beta=0$, for which all in-plane molecular orientations are equally populated. Despite this uniform angular averaging, the BSH fringes remain clearly discernible with appreciable contrast, whereas the spider-leg-like FSH structures are substantially smeared out. This contrasting response indicates that, under the considered sub-cycle driving conditions, the BSH structures are more robust against molecular-orientation averaging than the FSH structures.

Figures~\ref{fig7}(b)–(d) show the PMDs obtained for nonuniform angular distributions characterized by the degrees of alignment and orientation listed in Table~\ref{tab_costh}. In Fig.~\ref{fig7}(b), the ensemble is preferentially aligned along the laser polarization, with $\langle\cos^2\theta\rangle \approx 0.72$, but exhibits no net head-to-tail orientation, $\langle\cos\theta\rangle = 0$. Compared with the uniformly distributed ensemble in Fig.~\ref{fig7}(a), the BSH fringe contrast increases appreciably. Introducing a preferential head-to-tail orientation further enhances the fringe contrast, as shown in Fig.~\ref{fig7}(c), for which $\langle\cos^2\theta\rangle \approx 0.76$ and $\langle\cos\theta\rangle \approx 0.60$. Upon further increasing the alignment and orientation, as in Fig.~\ref{fig7}(d), with $\langle\cos^2\theta\rangle \approx 0.81$ and $\langle\cos\theta\rangle \approx 0.83$, the BSH fringe contrast approaches saturation. These results demonstrate that the BSH fringe visibility is sensitive to the angular distribution of the molecular ensemble and, for the distributions considered here, increases with preferential alignment and orientation before approaching saturation. Experimentally, polar molecules can be aligned and oriented using combinations of laser fields and weak static electric fields~\cite{Friedrich1999_JChemPhys, Stapelfeldt2003_RevModPhys, Hartelt2008_JChemPhys, Holmegaard2009_PRL, Ghafur2009_NatPhys, De2009_PRL}.

The persistence of the BSH structures in the angle-averaged PMDs suggests that they may provide access to molecular structural information even in experiments involving loosely aligned or broadly distributed molecular ensembles. More generally, these findings highlight the potential of sub-cycle laser fields for strong-field photoelectron holography, as the temporal confinement of the driving field enables the BSH contribution to remain distinguishable from competing interference structures after molecular-orientation averaging.

\subsection{Effect of sub-cycle pulse CEP on BSH interference}
\label{subsec5}

\begin{figure}[t]
\centering\includegraphics[width=1\columnwidth]{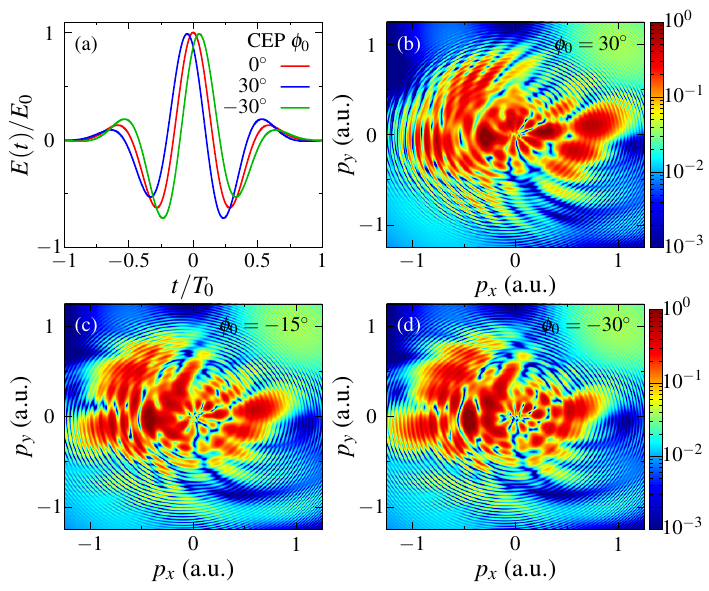}
\caption{(a) Temporal profiles of the laser electric field $\mathbf{E}(t)$ for different CEP values: $0^\circ$ (red), $30^\circ$ (blue), and $-30^\circ$ (green). The other laser parameters are identical to those used in Fig.~\ref{fig2}(a). (b)--(d) TDSE-simulated PMDs of CO for a molecular orientation angle of $\theta=45^\circ$ and CEP values of (b) $30^\circ$, (c) $-15^\circ$, and (d) $-30^\circ$.}
\label{fig8}
\end{figure}

Having established the robustness of the BSH interference fringes against molecular-orientation averaging, we now investigate their dependence on the CEP of the sub-cycle driving pulse. Owing to the sub-cycle duration of the pulse, its CEP plays an important role in determining the ionization and rescattering dynamics and, consequently, the visibility of different interference features in the PMD~\cite{Rajpoot2026_PRR}. Throughout the preceding analysis, the CEP was fixed at $\phi_0=0^\circ$. However, the experimental generation of CEP-stabilized sub-cycle laser pulses remains challenging. It is therefore important to examine how variations in the CEP affect the visibility and robustness of the BSH interference.

Figure~\ref{fig8}(a) shows the temporal profiles of the laser electric field for different CEP values around $\phi_0=0^\circ$, with all other laser parameters kept identical to those used in Fig.~\ref{fig2}(a). The corresponding TDSE-simulated PMDs of CO, for a fixed molecular orientation of $\theta=45^\circ$, are presented in Figs.~\ref{fig8}(b)--(d) for $\phi_0=30^\circ$, $-15^\circ$, and $-30^\circ$, respectively. Compared with the PMD at $\phi_0=0^\circ$ [Fig.~\ref{fig6}(c)], the BSH interference fringes remain clearly visible for $\phi_0=30^\circ$, as shown in Fig.~\ref{fig8}(b). In contrast, for negative CEP values, the visibility of the BSH fringes deteriorates more rapidly with increasing CEP deviation from zero. In these cases, the BSH structures become increasingly obscured by the spider-leg-like FSH interference features, as illustrated in Figs.~\ref{fig8}(c) and \ref{fig8}(d). These results indicate that, for the laser parameters considered here, positive CEP values favor the visibility of BSH interference. This suggests that choosing a CEP around $\phi_0\sim10^\circ$--$20^\circ$ may be advantageous for observing BSH structures under similar driving conditions.

To further investigate the robustness of the BSH interference against CEP variations, we calculate the CEP-averaged PMDs by uniformly averaging the CEP-resolved distributions over a specified CEP interval:
\be P_{\text{CEP}}(p_x,p_y) = \frac{1}{\Delta\phi_0} \int_{\phi_{0,\text{min}}}^{\phi_{0,\text{max}}} P_{\phi_0}(p_x,p_y) d\phi_0,
\label{eq_cep_average}
\ee
where $P_{\phi_0}(p_x,p_y)$ denotes the PMD calculated for a fixed CEP value $\phi_0$, and $\Delta\phi_0=\phi_{0,\text{max}}-\phi_{0,\text{min}}$ is the width of the CEP interval.

\begin{figure}[t]
\centering\includegraphics[width=1\columnwidth]{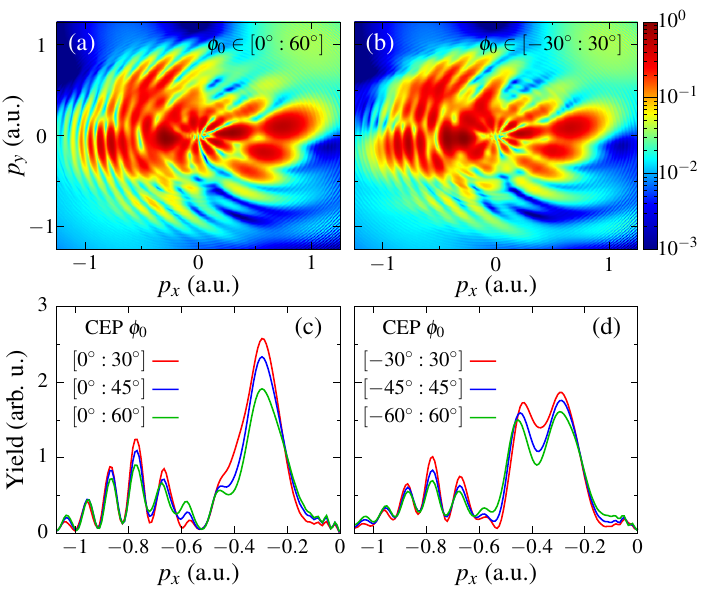}
\caption{CEP-averaged PMDs of CO for a molecular orientation angle of $\theta=45^\circ$, calculated using Eq.~\ref{eq_cep_average} over different CEP intervals: (a) $\phi_0\in[0^\circ,60^\circ]$ and (b) $\phi_0\in[-30^\circ,30^\circ]$. (c) and (d) Corresponding line profiles of the BSH interference structures extracted along $p_y=0$ a.u. for the positive CEP interval and the interval centered around $\phi_0=0^\circ$, respectively.}
\label{fig9}
\end{figure}

Figure~\ref{fig9} presents the resulting CEP-averaged PMDs for different CEP intervals. Interestingly, the BSH fringes remain discernible even after CEP averaging, although their contrast depends on the CEP interval considered. Figure~\ref{fig9}(a) shows the CEP-averaged PMD for the positive CEP range $\phi_0\in[0^\circ,60^\circ]$, in which the BSH interference fringes retain good contrast. For comparison, Fig.~\ref{fig9}(b) presents the CEP-averaged PMD over the range $\phi_0\in[-30^\circ,30^\circ]$. The BSH fringes remain visible in this case as well, although their contrast is slightly reduced compared with that obtained for the positive CEP range.

For a quantitative comparison, we extract line profiles of the BSH interference structures along $p_y=0$ a.u. from the CEP-averaged PMDs. Figures~\ref{fig9}(c) and \ref{fig9}(d) show the corresponding profiles for the positive CEP interval and the interval centered around $\phi_0=0^\circ$, respectively. The comparison confirms that the BSH fringes exhibit higher contrast for the positive CEP range. Furthermore, for both averaging intervals, the fringe contrast decreases as the CEP averaging range is extended. This reduction can be attributed to the superposition of PMDs with different CEP-dependent interference patterns, which progressively diminishes the contrast of the individual fringes.
Overall, these results demonstrate that the BSH interference remains observable over a finite range of CEP values and can survive CEP averaging, with its visibility depending on the CEP distribution. This robustness is an important consideration for assessing the experimental observability of BSH interference under realistic laser conditions. 

In experiments, the measured PMD is also affected by focal-volume averaging arising from the spatial variation of the laser intensity across the interaction volume. Such averaging can modify the relative contributions of different ionization pathways and may reduce the visibility of the interference structures. 

\begin{figure}[t]
\centering\includegraphics[width=1.0\columnwidth]{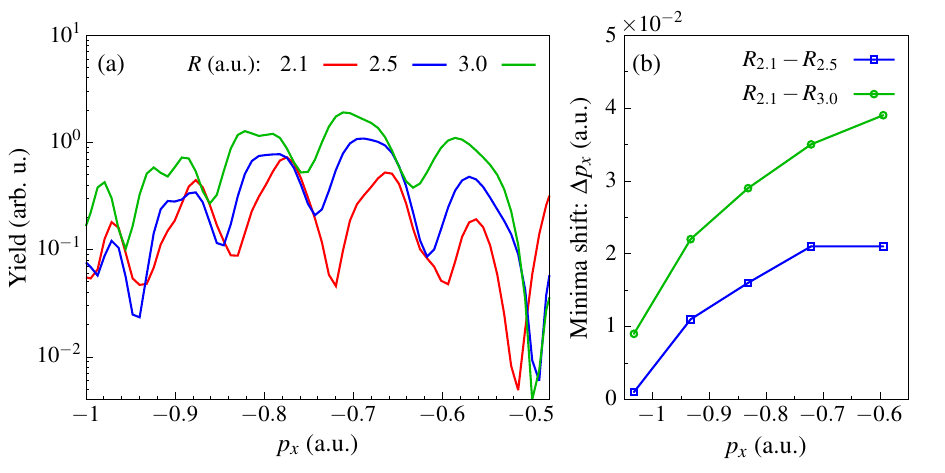}
\caption{ Influence of the internuclear separation $R$ on the BSH interference structures. (a) Line profiles of the BSH interference pattern extracted along $p_y=0$ a.u. from the TDSE-simulated PMDs of CO for a fixed molecular orientation of $\theta=45^\circ$ and different internuclear separations, $R = 2.1$, $2.5$, and $3.0$ a.u. (b) Relative shifts of the BSH interference minima for $R = 2.5$ a.u. (blue) and $3.0$ a.u. (green) with respect to the corresponding minima for $R = 2.1$ a.u.} 
\label{fig10}
\end{figure}

We finally investigate the sensitivity of the BSH interference pattern to the internuclear separation $R$. Figure~\ref{fig10}(a) presents line profiles extracted along $p_y=0$ a.u. from the TDSE-simulated PMDs of CO for $R=2.1$, $2.5$, and $3.0$ a.u., with the molecular orientation fixed at $\theta = 45^\circ$. To maintain a constant ionization potential of $I_p = 0.52$ a.u. across the different bond lengths, the soft-core parameter $a_0$ is adjusted accordingly. Specifically, $a_0 = 0.76$ and $0.79$ a.u. for $R = 2.5$ and $3.0$ a.u., respectively. As shown in Fig.~\ref{fig10}(a), the BSH fringe positions exhibit a clear dependence on $R$: with increasing internuclear separation, the interference minima shift toward larger values of the parallel-momentum magnitude, $|p_x|$. This shift reflects the interaction of the backward-scattered EWP with the molecular potential, indicating that the BSH interference pattern encodes information about the target structure. Consequently, the positions of the BSH minima can, in principle, be used to infer the internuclear separation.

The fringe spacing also exhibits a systematic dependence on $R$. For all three internuclear separations, the spacing between adjacent minima decreases with increasing $|p_x|$. The largest fringe spacing is obtained for $R=2.1$ a.u., while progressively smaller spacings are observed for $R=2.5$ and $3.0$ a.u. This behavior is further quantified in Fig.~\ref{fig10}(b), which shows the relative shifts of the BSH minima for $R=2.5$ and $3.0$ a.u. with respect to the corresponding minima for $R=2.1$ a.u. For both larger internuclear separations, the shift is larger at lower $|p_x|$ and gradually decreases toward higher parallel momenta. Furthermore, the shift is consistently larger for $R=3.0$ a.u. than for $R=2.5$ a.u. over the momentum range considered. These systematic variations demonstrate the sensitivity of the BSH interference pattern to the internuclear separation and highlight the structural information encoded in the interference fringes.

Taken together, the results presented above demonstrate three important characteristics of BSH interference in the CO molecule. First, the BSH fringe positions exhibit a clear and systematic dependence on the internuclear separation, indicating their sensitivity to the molecular structure. Second, unlike the FSH interference fringes, whose positions are strongly affected by molecular orientation, the BSH fringe positions remain nearly unchanged over the range of orientations investigated and persist after orientation averaging. Third, the BSH interference remains discernible over a finite range of CEP values and survives CEP averaging, although its contrast depends on the CEP interval considered. The combination of structural sensitivity and robustness against molecular-orientation and CEP averaging suggests that BSH interference generated by sub-cycle laser pulses could provide a promising route toward probing molecular structure under realistic experimental conditions. More broadly, these results demonstrate the potential of sub-cycle driving fields for isolating backward-scattering holographic contributions from other competing interference mechanisms in the PMD.

\section{Summary and Concluding Remarks}
\label{sec4}

We have systematically investigated photoelectron holographic interference in the CO molecule driven by a half-cycle laser pulse using 2D-TDSE simulations and a classical-trajectory-based analysis. The sub-cycle temporal confinement of the driving field enables the FSH and BSH interference structures in the PMDs to be more clearly distinguished. We first examined the dependence of the FSH interference on the molecular orientation. The FSH fringes exhibit a pronounced shift as the molecular orientation angle is varied relative to the laser polarization direction. Analysis of the electron density near the ionization region indicates that this shift is associated with the orientation-dependent transverse displacement of the ionizing EWP, consistent with previous studies~\cite{Tan2023_PRA,Meckel2014_NatPhys,Liu2016_PRL}. Furthermore, the heteronuclear nature of CO leads to different transverse displacements, and consequently different FSH fringe shifts, when ionization occurs predominantly from the C or O side of the molecule.

We then identified the origin of the curved interference structures in the negative parallel-momentum region of the PMD as BSH interference. A classical-trajectory-based analysis of the possible intracycle interference pathways shows that competing FSH and TDS contributions involving O-side ionization are strongly suppressed under the present field configuration. In contrast, the BSH pathway involves EWPs predominantly ionized from the C side during the strong QCs surrounding the peak of the sub-cycle pulse. The resulting coherent superposition of direct and backward-scattered EWPs gives rise to the characteristic right-curved BSH fringes observed in the PMDs.

An important feature of the BSH interference structures is their robustness with respect to molecular orientation. While the positions of the FSH fringes change appreciably with molecular orientation, the positions of the BSH fringes remain nearly unchanged over the range of orientations investigated. Moreover, the BSH structures remain clearly visible after orientation averaging, whereas the spider-leg-like FSH structures are substantially smeared out. This robustness suggests that BSH interference may provide a useful probe of molecular structure in experiments where precise control of the molecular orientation is challenging.

We further investigated the influence of the CEP of the sub-cycle driving pulse on the BSH interference. The CEP-resolved PMDs show that the visibility of the BSH fringes depends on the CEP, with positive CEP values favoring their visibility under the laser conditions considered. In particular, the BSH structures remain clearly discernible for CEP values around $10^\circ$--$20^\circ$, whereas for negative CEP values they become increasingly obscured by the FSH interference features. Importantly, the BSH fringes also persist after CEP averaging over the intervals considered, although their contrast depends on the CEP distribution. These findings demonstrate that BSH interference can remain observable even in the presence of finite CEP variations, an important consideration for its experimental observation.

Finally, we demonstrated that the BSH interference contains information about the molecular structure. By varying the internuclear separation $R$ while keeping the molecular orientation fixed, we found a systematic shift of the BSH interference minima toward larger parallel momentum magnitudes with increasing $R$. The fringe spacing also changes systematically with the internuclear separation. These results demonstrate the sensitivity of the BSH interference pattern to the molecular geometry and indicate that the positions of the BSH minima can, in principle, serve as observables for extracting structural information.

Overall, the combination of structural sensitivity and robustness against molecular-orientation and CEP averaging highlights the potential of backward-scattering photoelectron holography as a probe of molecular structure. The present results further demonstrate that sub-cycle driving fields can help isolate BSH interference from competing direct-electron and non-holographic contributions, providing a clear framework for studying backward-scattering holography in molecules. Polar molecules offer a particularly interesting platform for investigating site-selective ionization and for exploring the interplay between the molecular dipole and the ionic Coulomb potential. In this context, a recent study demonstrated clear orientation-dependent asymmetries in the PMD of a heteronuclear molecule~\cite{Haertelt2026_PRL}. The present analysis opens the possibility of applying sub-cycle pulse enabled backward-scattering holography to investigate site-specific ionization pathways and, ultimately, their connection to charge migration in ultrafast processes. Extending the present analysis toward these aspects will therefore be an important direction for future work.



\section*{Acknowledgments} Authors would like to acknowledge the financial support from the Ministry of Education, Culture, Sports, Science and Technology of Japan (MEXT) through Grants-in-Aid under grant no. 21H01850, and the MEXT Quantum Leap Flagship Program (Q-LEAP) (grant no. JP-MXS0118068681). This project was supported by the RIKEN TRIP initiative.


%

\end{document}